\documentclass[english,pra,showpacs,showkeys,tightenlines,secnumarabic,11pt]{revtex4}
\usepackage[T1]{fontenc}
\usepackage[latin1]{inputenc}
\usepackage{amsmath}
\usepackage{graphicx}
\usepackage{amssymb}
\usepackage{epsfig}
\usepackage{graphics}
\usepackage[mathscr]{euscript}
\usepackage{psfrag}
\usepackage{pstricks}
\usepackage{pst-node}
\usepackage{babel}

\begin{document}
\title{Double parton scattering, diffraction and effective cross section}
\author{D. Treleani}
\email{daniele.treleani@ts.infn.it}
\affiliation{ Dipartimento di Fisica Teorica dell'Universit\`a di Trieste,
INFN, Sezione di Trieste and ICTP\\ Strada Costiera 11, Miramare-Grignano,
I-34014 Trieste, Italy.}
\begin{abstract}
The rates of multiparton collisions in high energy hadronic interactions provide information on the
typical transverse distances between partons in the hadron
structure. The different configurations of the hadron in
transverse space are, on the other hand, at the origin of hadron
diffraction. The relation between the two phenomena is exploted in
an eikonal model of hadronic interactions.
\end{abstract}
\pacs{11.80.Gw, 11.80.La, 13.60.Hb, 13.85.Hd} \keywords{High energy hadronic interactions, Multiple Scattering}
 \maketitle
\section{Introduction}

The high parton luminosity and the proximity with the initial runs
at the LHC have generated a renewed interest in multiple parton
interactions\cite{DESY}. The rate of multiple parton interactions
is in fact estimated to be rather large at the LHC\cite{Catani:2000jh}\cite{Del Fabbro:2002pw} giving rise to
different effects, among others a substantial increase of the
estimated background in channels of interest for the search of new
physics (e.g. for Higgs discovery\cite{Del Fabbro:1999tf}).
Unfortunately the process cannot be estimated in a straightforward
way, due to the lack of knowledge of the non-perturbative input,
namely the multi-parton correlations of the hadron structure.
Existing experimental results on multiple parton
interactions\cite{Akesson:1986iv}\cite{Abe:1997bp}\cite{Abe:1997xk}
show in fact that the simplest implementation of the phenomenon,
where multi-parton correlations are neglected, is unable to
reproduce the data\cite{Calucci:1999yz}\cite{Frankfurt:2004kn}.

The unexpected feature of multiple parton interactions, emerged
from the experimental studies of the phenomenon, is represented by
the small value of the "effective cross section", the scale factor
characterizing double parton collisions.

The simplest possibility for multiparton interactions is a
Poissonian distribution, at a fixed hadronic impact parameter\cite{Capella:1986cm}\cite{Ametller:1987ru}\cite{Gotsman:1990de}\cite{Corsetti:1996wg}.
The Poissonian at fixed impact parameter takes into account only
the most obvious correlation between partons, namely that all
partons must be localized inside the volume occupied by the
hadron. An argument in favour of this simplifying attitude is that
other sources of correlations, e.g. conservation laws, are diluted
at small $x$ (where multiple parton collisions are a sizable
effect) by the large parton population.

\vskip .25in
\par   {\bf Inclusive and exclusive cross sections in the Poissonian model}
\vskip .15in

In the Poissonian model one introduces the three dimensional
parton density $\Gamma(x,b)$, representing the average number of
partons with a given momentum fraction $x$ and transverse
coordinate $b$ (the dependence on flavor and on the resolution of
the process is implicit) and one makes the simplifying assumption
that the dependence on the transverse and longitudinal degrees of
freedom can be factorized as $\Gamma(x,b)=G(x)f(b)$, with $G(x)$
the usual parton distribution function and $f(b)$ a function
normalized to one and representing the distributions of partons in
transverse space. The inclusive cross section for large $p_t$
parton production may hence be expressed as

\begin{eqnarray}
\sigma_S&=&\int_{p_t^c}G(x)\hat{\sigma}(x,x')G(x') dxdx'\nonumber\\&=&\int_{p_t^c}G(x)f(b) \hat{\sigma}(x,x')G(x')f(b-\beta)
         d^2 b dxdx'd^2\beta
\end{eqnarray}

\noindent where $\hat{\sigma}(x,x')$ is the parton-parton cross
section integrated with the cutoff $p_t^c$, introduced to
distinguish hard and soft parton collisions, $\beta$ is the impact
parameter of the hadronic interaction and $b$ is the transverse
coordinate of the two colliding partons (the relative transverse
distance between the interacting partons is neglected, being here
compared with the much larger hadronic dimension).

\noindent Disregarding the effects of correlations in the multi-parton distributions, the inclusive cross section for a double parton
scattering $\sigma_D$ is analogously expressed by

\begin{eqnarray}
\sigma_D&=&\frac{1}{ 2!}\int_{p_t^c}G(x_1)f(b_1)\hat{\sigma}(x_1,x_1')G(x_1')f(b_1-\beta)d^2b_1dx_1dx_1'\times\nonumber\\
&&\qquad\qquad\times
         G(x_2)f(b_2)\hat{\sigma}(x_2,x_2') G(x_2')f(b_2-\beta)
         d^2b_2dx_2dx_2'd^2\beta\nonumber\\
         &=&\frac{1}{ 2!}\int\Big(\int_{p_t^c} G(x)f(b)\hat{\sigma}(x,x')G(x')f(b-\beta)d^2bdxdx'\Big)^2d^2\beta\nonumber\\
         &=&\frac{1}{ 2}\frac{\sigma_S^2}{ \sigma_{eff}}
\end{eqnarray}

\noindent where the scale factor $\sigma_{eff}^{-1}=\int
d^2\beta\bigl[F(\beta)\bigr]^2$, with $F(\beta)=\int
f(b)f(b-\beta)d^2b$, has been introduced\cite{Paver:1982yp}. As
$\sigma_D$ is obtained by multiplying $\sigma_S$ by the ratio
$\sigma_S/\sigma_{eff}$, the scale factor $\sigma_{eff}$
represents the value of $\sigma_S$ where the inclusive rate of
double collisions becomes as large as the inclusive rate of single
collisions (apart from the factor $1/2$ due to the identity of the
two interactions). Similarly the Nth parton scattering inclusive
cross section $\sigma_N$ is given by

\begin{eqnarray}
\sigma_N&=&\int\frac{1}{ N!}\Big(\int_{p_t^c}G(x)f(b)\hat{\sigma}(x,x')G(x')f(b-\beta)d^2bdxdx'\Big)^Nd^2\beta\nonumber\\
&=&\int\frac{1}{ N!}\bigl(\sigma_SF(\beta)\bigr)^Nd^2\beta
\end{eqnarray}

In the simplest uncorrelated case one may easily give explicit
expressions also for the {\it exclusive} hard cross sections,
namely for the various contributions to the inelastic cross
section due to the different multiparton scattering processes. The
integrand $\frac{1}{ N!}\bigl(\sigma_SF(\beta)\bigr)^N$ is in fact
dimensionless and, once normalized, it may be understood as the
probability for a Nth parton collision process.

\noindent The hard cross section $\sigma_{hard}$, namely the
contribution to the inelastic cross section given by all events
with {\it at least} one parton collision with momentum transfer
greater than the cutoff $p_t^c$, is hence expressed by

\begin{eqnarray}
\sigma_{hard}
 =\sum_{N=1}^{\infty}\int d^2\beta\frac{\bigl(\sigma_SF(\beta)\bigr)^N}{ N!}
 e^{-\sigma_SF(\beta)}=\int d^2\beta\Bigl[1-e^{-\sigma_SF(\beta)}\Bigr]
\end{eqnarray}

\noindent
and

\begin{equation}
\frac{\bigl(\sigma_SF(\beta)\bigr)^N}{ N!}
 e^{-\sigma_SF(\beta)}={\rm P}_N(\beta)
\end{equation}

\noindent represents the probability of having N parton collisions in a
hadronic interaction at impact parameter $\beta$. The relation
with the inelastic cross section is

\begin{equation}
\sigma_{inel}=\sigma_{soft}+\sigma_{hard}
\end{equation}

\noindent with $\sigma_{soft}$ the soft contribution to the
inelastic cross section $\sigma_{inel}$, the two contributions
$\sigma_{soft}$ and $\sigma_{hard}$ being defined through the
cutoff in the momentum exchanged between partons, $p_t^c$. Notice
that, differently from the case of the inclusive cross sections,
which are divergent for $p_t^c\to0$, both $\sigma_{hard}$ and all
exclusive contributions to $\sigma_{hard}$, with a given number of
parton collisions, are finite in the infrared limit.

\vskip .25in
\par   {\bf Dispersion and effective cross section}
\vskip .15in

The many-parton inclusive cross sections
are proportional to the moments of the distribution in multiple
parton collisions.

\noindent In particular the single and the double parton inclusive
scattering cross sections are proportional to the average number of
parton scatterings and to the dispersion of the distribution in
the number of collisions. The average number of parton scatterings
is in fact given by:

\begin{eqnarray}
\langle N\rangle\sigma_{hard}=\int d^2\beta\sum_{N=1}^{\infty}
\frac{N\bigl[\sigma_SF(\beta)\bigr]^N}{ N!}
 e^{-\sigma_SF(\beta)}
=\int d^2\beta \sigma_SF(\beta)=\sigma_S
\end{eqnarray}

\noindent
while for the dispersion one obtains:

\begin{eqnarray}
\frac{\langle N(N-1)\rangle}{2}\sigma_{hard}
&=&\frac{1}{2}\int d^2\beta \sum_{N=2}^{\infty}
\frac{N(N-1)\bigl[\sigma_SF(\beta)\bigr]^N}{ N!}
 e^{-\sigma_SF(\beta)}\nonumber\\
&=&\frac{1}{2}\int d^2\beta \bigl[\sigma_SF(\beta)\bigr]^2=\sigma_D
\end{eqnarray}

\noindent The relations between $\sigma_S$ and $\langle N\rangle$
and between $\sigma_D$ and $\langle N(N-1)\rangle$ are not
peculiar of the simplest Poissonian distribution. Their validity is
indeed much more
general\cite{Calucci:1991qq}\cite{Calucci:1999yz}. The same
relations can in fact be obtain also when considering the most
general case of multiparton distributions, namely including all
possible multi-parton correlations (in particular the correlations
induced by conservation laws). On the other hand, the direct
link, between inclusive cross sections and moments of the
distribution in multiplicity of collisions, gets spoiled when
taking into account connected multiparton interactions, namely
$3\to3$ etc. parton collision processes, which nevertheless should
not give rise to major effects in $pp$ collisions even at the LHC
energy. Limiting the discussion to the case of hadronic
interaction, namely excluding hadron-nucleus and
nucleus-nucleus collisions, one may hence write

\begin{equation}
\langle N\rangle\sigma_{hard}=\sigma_S\quad {\rm and}\quad\frac{1}{2} \langle
N(N-1)\rangle\sigma_{hard}=\sigma_D
\end{equation}

\noindent The effective cross section $\sigma_{eff}$ is therefore
linked to the dispersion $\langle N(N-1)\rangle$ and to the
average $\langle N\rangle$ by the relation:

\begin{equation}
\langle N(N-1)\rangle=\langle N\rangle^2\frac{\sigma_{hard}}{\sigma_{eff}}
\end{equation}

\noindent which implies that, if one had a Poissonian distribution
of multiple parton collisions also after integration over the
impact parameter $\beta$, one would have
$\sigma_{eff}=\sigma_{hard}$. Even in the simplest
case, the distribution is however Poissonian before integration on
the impact parameter. The final distribution in the number of
parton collisions will hence have a larger dispersion as compared
with the Poissonian, which implies $\sigma_{eff}$ smaller
than $\sigma_{hard}$.

\noindent The actual values of $\sigma_{eff}$ and of
$\sigma_{hard}$ depend on the functional form used for $F(\beta)$.
For $F(\beta)={\rm exp}(-\beta^2/R^2)/\pi R^2$ one obtains:

\begin{equation}
\sigma_{hard}=2\pi R^2\bigl[\gamma+{\rm ln}\kappa+E_1(\kappa)\bigr]
\end{equation}

\par\noindent
where $\gamma=0.5772\dots$ is Euler's constant,
$\kappa=\sigma_S/(2\pi R^2)$ and $E_1(x)$ is the exponential
integral.

\par\noindent
For $\kappa$ small $\sigma_{hard}\to 2\pi R^2\kappa=\sigma_S$,
while for $\kappa$ large (namely $\sigma_S\to \infty$) one obtains
$\sigma_{hard}\to2\pi R^2\bigl(\gamma+{\rm}ln\kappa\big)$. Here
$\sigma_{eff}=2\pi R^2$. The value of $\sigma_{hard}$ is therefore
proportional to the value of $\sigma_{eff}$, the proportionality
factor being slightly dependent on energy and cutoff. Sensible
values of the hadron-hadron c.m. energy and of the cutoff give
values of $\sigma_{hard}$ $30-40\%$ larger than $\sigma_{eff}$.
Asymptotically (at high energy and fixed momentum cutoff) one
expects $\sigma_{hard}\approx\sigma_{inel}$. The simplest
expectation is hence that $\sigma_{eff}$ should not be much
smaller than $\sigma_{inel}$.

On the contrary the experimental
indication\cite{Akesson:1986iv}\cite{Abe:1997bp}\cite{Abe:1997xk}
is that $\sigma_{eff}$ may be more than a factor three smaller
than $\sigma_{inel}$.  Although one may obtain a value of
$\sigma_{eff}$ sizably smaller than $\sigma_{inel}$ choosing
appropriate analytic forms for $F(\beta)$\cite{Sjostrand:2006za},
sensible choices of $F(\beta)$ give results qualitatively similar
to the simplest gaussian case\cite{Calucci:1999yz}.

An obvious possibility to increase the dispersion of the
distribution in the number of collisions, namely to obtain smaller
values of $\sigma_{eff}$, is to consider a different distribution
in the number of collisions at a fixed impact parameter, which
evidently implies a non secondary role of correlations in the
multiparton distribution. Including correlations one may hence
easily obtain a large dispersion in the final distribution of
multiple parton interactions and, correspondingly, a small value
of $\sigma_{eff}$\cite{Del Fabbro:2000ds}.

A source of correlation for the parton population is the
fluctuation of the hadron in its transverse dimension, which is a
phenomenon related directly to hadronic diffraction. Although the
diffractive cross section is expected to be small at high
energies, recent estimates indicate sizably large effects of
diffraction in $pp$ collisions also at the
LHC\cite{Gotsman:1993ux}\cite{Gotsman:1999ri}\cite{Gotsman:2000gb}\cite{Gotsman:2007pn}.

The purpose of the present paper is to study the link between diffraction and multiparton collisions. The dispersion in the collisions distribution, and hence of the expected value of the effective cross
section, will thus be derived in the simplest multichannel eikonal
model of high energy hadronic interactions, capable of reproducing
the total, elastic, inelastic, single and double diffractive cross
sections in high energy $pp$ collisions.

To make the argument self sufficient, the multi-channel eikonal
model will be discussed in detail in the next section. The cross
sections of inclusive and exclusive hard processes will hence be
derived in the following paragraph. The effective cross section,
at TeVatron and LHC energies, will be finally estimated in the
simplest two-channel case, using as input the parameters fitted to
reproduce the available information on soft cross sections in
high energy $pp$ collisions.

\vskip.2in

\section{The multi-channel eikonal model}

Let's represent with $\psi_{\mu}$ the initial hadron and its
diffractive states and with $\phi_{\eta}$
the eigenstates of the forward $T$ matrix scattering operator and
let's normalize the imaginary part of the nucleon-nucleon forward
$T$ matrix operator to the total cross section of the different
hadronic channels $\mu$ $\nu$, $[\sigma_{tot}]_{\mu\nu}$:

\begin{equation}{\rm Im}\langle\psi_{\mu} \psi_{\nu} |T|\psi_{\mu}\psi_{\nu}\rangle=[\sigma_{tot}]_{\mu\nu}\end{equation}

\noindent here and in the following the indices ${\mu}$ and ${\nu}$ will refer either
to the hadron or to its various diffractive states, while the indices $\eta$ and $\rho$ to the eigenstates of the forward $T$ matrix scattering operator. One
may hence write:

\begin{equation}{\rm Im}\langle\psi_{\mu} \psi_{\nu} |T|\psi_{\mu}\psi_{\nu}\rangle=\sum_{\eta\rho}|a_{\mu\eta}|^2|a_{\nu\rho}|^2{\rm Im}T_{\eta\rho}\end{equation}

\noindent where $T_{\eta\rho}=\langle\phi_\eta \phi_\rho
|T|\phi_\eta\phi_\rho\rangle$ and $a_{\mu\eta}=\langle\psi_\mu
|\phi_\eta\rangle$.

\noindent Let's denote with $t_{\eta\rho}$ the one-Pomeron
exchange forward scattering amplitude of the states $\phi_\eta$
and $\phi_\rho$, with
imaginary part normalized to the cross
section $\sigma_{\eta\rho}$.

\noindent In the case of interest large numbers of Pomerons may be
exchanged.

\noindent Assuming independency for the different
exchanges, one may write:

\begin{equation}T_{\eta\rho}=\sum_{k=1}^{\infty}\langle\phi_\eta \phi_\rho |t_k|\phi_\eta\phi_\rho\rangle
\qquad{\rm where}\qquad\frac{{\rm i}}{2}\langle\phi_\eta
\phi_\rho|t_k|\phi_\eta\phi_\rho\rangle= \int
d^2b\frac{1}{k!}\biggl(\frac{{\rm i}t_{\eta\rho}(b)}{
2}\biggr)^k\end{equation}

\noindent and

\begin{equation} \frac{\rm i}{ 2}T_{\eta\rho}=\int d^2b\sum_{k=1}^{\infty}\frac{1}{k!}\biggl(\frac{{\rm i}t_{\eta\rho}(b)}{
2}\biggr)^k=\int
d^2b\Bigl\{{\rm e}^{\frac{{\rm i}t_{\eta\rho}(b)}{
2}}-1\Bigr\} \end{equation}

\noindent The total cross section between the eigenstates of the
$T$ matrix $\phi_{\eta}$ and $\phi_{\rho}$ is obtained through the
optical theorem:

\begin{equation}{\sigma^{\eta\rho}_{tot}=2{\rm Re}\int d^2b\Bigl\{ 1-{\rm e}^{\frac{{\rm i}t_{\eta\rho}(b)}{
2}} \Bigr\}}\end{equation}

\noindent which may be also expressed as a sum of multiple collisions
amplitudes:

\begin{equation}\sigma^{\eta\rho}_{tot}=2{\rm Re}\int d^2b\sum_{k=1}^{\infty}\frac{1}{k!}\biggr(\frac{{\rm i}t_{\eta\rho}(b)}{
2}\biggr)^k{\rm
e}^{-\frac{{\rm i}t_{\eta\rho}}{
2}}\end{equation}

\noindent the total cross sections between the hadronic states
$\psi_{\mu}$ and $\psi_{\nu}$ is hence given by

\begin{equation}[\sigma_{tot}]_{\mu\nu}=\sum_{\eta\rho}|a_{\mu\eta}|^2|a_{\nu\rho}|^2\sigma^{\eta\rho}_{tot}\end{equation}

\par
The different (leading) contributions to the cross section due to
the various elastic, diffractive and production channels are
readily derived\cite{Abramovsky:1973fm}. Let's first consider the
contributions to the cuts due to the scattering between the
eigenstates of the the $T$ matrix  $\phi_{\eta}$ and
$\phi_{\rho}$, $T_{\eta\rho}$. Each $t_{\eta\rho}$ represents a
single Pomeron exchange; let's denote with $D_{k,m}^{\eta\rho}$
the contribution to the inelastic cross section where $k$ Pomerons
are exchanged and $m$ Pomerons are cut, $k\geq m$. From Eq.(15)
one obtains:

\begin{equation}D_{k,m}^{\eta\rho}= \int d^2b\frac{1}{k!}\binom{k}{ m}
                      \sum_{l=0}^{k-m}\binom{k-m}{ l}\biggl[\frac{{\rm i}t_{\eta\rho}(b)}{2}\biggr]^l\biggl[\biggl(\frac{{\rm i}t_{\eta\rho}(b)}{
                      2}\biggr)^*\biggr]^{k-m-l}\times\sigma_{\eta\rho}(b)^m\end{equation}

\vskip.1in - a factor ${\rm i}t_{\eta\rho}(b)/ 2$ is associated to each Pomeron on
the left hand side of the cut,

\vskip.1in - a factor $({\rm i}t_{\eta\rho}(b)/ 2)^*$ is
associated to each Pomeron on the right hand side of the cut

\vskip.1in -
a factor $2{\rm Im}({\rm i}t_{\eta\rho}(b)/2)=\sigma_{\eta\rho}(b)$ is associated to
each cut-Pomeron.

\par\noindent
The sum over $l$ represents all possibilities, namely $l$ Pomerons
on the left hand side of the cut and $k-m-l$ Pomerons on the right hand side
of the cut.

\par\noindent
The sum over $l$ can be performed giving:

\begin{eqnarray}
D_{k,m}^{\eta\rho}&=&\int d^2b \frac{1}{k!}\binom{k}{ m}
 \biggl[
 \frac{{\rm i}t_{\eta\rho}(b)}{2}+\biggl(\frac{{\rm i}t_{\eta\rho}(b)}{2}\biggr)^*
 \biggr]^{k-m}\times \sigma_{\eta\rho}(b)^m\nonumber \\
&=&\int d^2b
\frac{1}{k!}\binom{k}{ m}
\bigl(
-\sigma_{\eta\rho}(b)
\bigr)^{k-m}
\times\sigma_{\eta\rho}(b)^m
\end{eqnarray}

\noindent where the relation ${\rm i}t_{\eta\rho}(b)/2+({\rm i}t_{\eta\rho}(b)/2)^*=-\sigma_{\eta\rho}(b)$ has been used.
\par\noindent The contribution to the cross section $\sigma_m^{\eta\rho}$, corresponding to $m$
cut-Pomerons, is obtained by summing over all elastic scatterings
in $D_{k,m}^{\eta\rho}$:

\begin{eqnarray}
\sigma_m^{\eta\rho}=\sum_{k=m}^{\infty}D_{k,m}^{\eta\rho}&=&\int d^2b\sum_{k=m}^{\infty}
\frac{1}{k!}\frac{k!}{(k-m)!m!}
                        \bigl(-\sigma_{\eta\rho}(b)\bigr)^{k-m}\sigma_{\eta\rho}(b)^m\cr
                 &=&\int d^2b\frac{1}{m!}\bigl(\sigma_{\eta\rho}(b)\bigr)^m
                            {\rm
e}^{-\sigma_{\eta\rho}(b)} \end{eqnarray}

\par\noindent
The contribution to the inelastic cross sections due to the
scattering between the eigenstates of the the $T$ matrix
$\phi_{\eta}$ and $\phi_{\rho}$ is the result of summing all
$\sigma_m^{\eta\rho}$'s:

\begin{equation}
\sigma_{in}^{\eta\rho}=\sum_{m=1}^{\infty}\sigma_m^{\eta\rho}=\int d^2b\sum_{m=1}^{\infty}\frac{1}{m!}\bigl(\sigma_{\eta\rho}(b)\bigr)^m
                            {\rm
e}^{-\sigma_{\eta\rho}(b)}=\int d^2b\Bigl\{1-{\rm
e}^{-\sigma_{\eta\rho}(b)}\Bigr\} \end{equation}

\noindent The inelastic cross sections of the hadronic states
$\psi_{\mu}$ and $\psi_{\nu}$ is hence given by

\begin{equation}[\sigma_{in}]_{\mu\nu}=\sum_{\eta\rho}|a_{\mu\eta}|^2|a_{\nu\rho}|^2\sigma_{in}^{\eta\rho}\end{equation}

The elastic and diffractive cuts of the scattering amplitude of
the hadronic states $\psi_{\mu}$ and $\psi_{\nu}$ are obtained
from the elastic cuts of the scattering between the eigenstates of
the $T$ matrix $\phi_{\eta}$ and $\phi_{\rho}$ (of course the only
possibilities for the eigenstates of the $T$ matrix is to be
either absorbed or to scatter elastically).

\par\noindent
One my obtain the elastic cross section of the scattering of
$\phi_{\eta}$ and $\phi_{\rho}$ working out all elastic cuts
between the exchanged Pomerons. Let's introduce
$E_{k}^{\eta\rho}$, the contribution to the elastic scattering
cross section of the eigenstates $\phi_{\eta}$ and $\phi_{\rho}$
with $k$ Pomerons exchanged. Obviously elastic cuts are possible
only for $k\geq 2$:

\begin{eqnarray}E_{k}^{\eta\rho}&=&\int d^2b \frac{1}{k!}\biggl\{
                      \sum_{l=1}^{k-1}\binom{k}{ l}\biggl[\frac{{\rm i}t_{\eta\rho}(b)}{2}\biggr]^l
                      \biggl[\biggl(\frac{{\rm i}t_{\eta\rho}(b)}{2}\biggr)^*\biggr]^{k-l}\biggr\}\cr
                      &=&\int d^2b \frac{1}{k!}\biggl\{
                      \biggl[\frac{{\rm i}t_{\eta\rho}(b)}{2}+
                     \biggl(\frac{{\rm i}t_{\eta\rho}(b)}{2}\biggr)^*\biggr]^{k}-\biggl[\biggl(\frac{{\rm i}t_{\eta\rho}(b)}{2}
                     \biggr)^*\biggr]^k-\biggl[\biggl(\frac{{\rm i}t_{\eta\rho}(b)}{2}\biggr)\biggr]^k\biggr\}\cr
                      &=&\int d^2b \frac{1}{k!}\biggl\{\bigl(-\sigma_{\eta\rho}(b)\bigr)^k-2\times{\rm Re}\biggl(-\frac{\sigma_{\eta\rho}(b)}{
                      2}(1-{\rm i}\alpha_{\eta\rho})\biggr)^k\biggr\}
                      \end{eqnarray}

\par\noindent The contribution due to the elastic scattering between the
eigenstates of the the $T$ matrix $\phi_{\eta}$ and $\phi_{\rho}$
is obtained by summing over all elastic cuts:

\begin{eqnarray}
\sigma_{el}^{\eta\rho}=\sum_{k=2}^{\infty}E_{k}^{\eta\rho}&=&\int
d^2b \Bigl\{{\rm
e}^{-\sigma_{\eta\rho}(b)}-1+\bigl(-\sigma_{\eta\rho}(b)\bigl)\cr&&\qquad
\qquad-2\times{\rm Re}\Bigl[{\rm e}^{-\frac{\sigma_{\eta\rho}(b)}{
2}(1-{\rm
i}\alpha_{\eta\rho})}-1+\Bigr(-\frac{\sigma_{\eta\rho}(b)}{
2}(1-{\rm i}\alpha_{\eta\rho})\Bigl)\Bigr]\Bigr\}\cr &=&\int
d^2b\Bigl|1-{\rm e}^{-\frac{\sigma_{\eta\rho}(b)}{ 2}(1-{\rm
i}\alpha_{\eta\rho})}\Bigr|^2
                      \end{eqnarray}

\par\noindent The total cross section of the scattering between the
eigenstates of the the $T$ matrix $\phi_{\eta}$ and $\phi_{\rho}$
results from the sum of the elastic and inelastic contributions:

\begin{eqnarray}\sigma_{el}^{\eta\rho}+\sigma_{in}^{\eta\rho}&=&\int d^2b\Bigl\{1-2\times{\rm
Re}\Bigl[{\rm e}^{-\frac{\sigma_{\eta\rho}(b)}{ 2}(1-{\rm
i}\alpha_{\eta\rho})}\Bigr]+{\rm
e}^{-{\sigma_{\eta\rho}(b)}}+1-{\rm
e}^{-{\sigma_{\eta\rho}(b)}}\Bigr\}\cr &=&2\int d^2b\Bigl\{{\rm
Re}\Bigl[ 1-{\rm e}^{-\frac{\sigma_{\eta\rho}(b)}{ 2}(1-{\rm
i}\alpha_{\eta\rho})}\Bigr] \Bigr\}=\sigma^{\eta\rho}_{tot}
\end{eqnarray}

\par\noindent which is precisely the expression for $\sigma^{\eta\rho}_{tot}$ in
Eq.16.

\noindent The total and inelastic cross sections between the
hadronic states $\psi_{\mu}$ and $\psi_{\nu}$ are hence given by

\begin{equation}[\sigma_{tot}]_{\mu\nu}=\sum_{\eta\rho}|a_{\mu\eta}|^2|a_{\nu\rho}|^2\sigma^{\eta\rho}_{tot}\end{equation}

\begin{equation}[\sigma_{in}]_{\mu\nu}=\sum_{\eta\rho}|a_{\mu\eta}|^2|a_{\nu\rho}|^2\sigma_{in}^{\eta\rho}\end{equation}

\noindent The sum of the elastic ($el$), single diffractive ($SD$)
and double diffractive ($DD$) cross sections in $hh$ collisions is
easily obtained:

\begin{eqnarray}\sigma_{el}+\sigma_{SD}+\sigma_{DD}
=\sum_{\eta\rho\gamma\sigma}\langle\psi_{h} \psi_{h}
|\phi_{\eta}\phi_{\rho}\rangle&\times& \langle\phi_\eta \phi_\rho
|T|\phi_\eta\phi_\rho\rangle \langle\phi_{\eta} \phi_{\rho}
|\sum_{\mu\nu}|\psi_{\mu}\psi_{\nu}\rangle\langle\psi_{\mu}
\psi_{\nu}|\cr&&\qquad\times|\phi_{\gamma}\phi_{\sigma}\rangle
\langle\phi_\gamma \phi_\sigma
|T^{\dag}|\phi_\gamma\phi_\sigma\rangle \langle\phi_{\gamma}
\phi_{\sigma} |\psi_{h}\psi_{h}\rangle\end{eqnarray}

\noindent The sum over the final states $\mu\nu$ gives one for
completeness and one is left with the product

\begin{equation}\langle\phi_{\eta}
\phi_{\rho}|\phi_{\gamma}\phi_{\sigma}\rangle=\delta_{\eta\gamma}\delta_{\rho\sigma}\end{equation}

\noindent One hence obtains

\begin{eqnarray}\sigma_{el}+\sigma_{SD}+\sigma_{DD}&=&
\sum_{\eta\rho}\big|\langle\psi_{h} \psi_{h}
|\phi_{\eta}\phi_{\rho}\rangle\big|^2\big|\langle\phi_\eta
\phi_\rho
|T|\phi_\eta\phi_\rho\rangle\big|^2\cr&=&\sum_{\eta\rho}|a_{h\eta}|^2|a_{h\rho}|^2\sigma_{el}^{\eta\rho}\end{eqnarray}

\noindent since

\begin{equation}\langle\psi_{h} \psi_{h}
|\phi_{\eta}\phi_{\rho}\rangle=a_{h\eta}a_{h\rho}\qquad{\rm
and}\qquad \big|\langle\phi_\eta \phi_\rho
|T|\phi_\eta\phi_\rho\rangle\big|^2=\sigma_{el}^{\eta\rho}\end{equation}

\noindent and summing the inelastic cross section

\begin{eqnarray}\sigma_{el}+\sigma_{SD}+\sigma_{DD}+\sigma_{in}&=&
\sum_{\eta\rho}|a_{h\eta}|^2|a_{h\rho}|^2\bigl(\sigma_{el}^{\eta\rho}+\sigma_{in}^{\eta\rho}\bigr)\cr
&=&\sum_{\eta\rho}|a_{h\eta}|^2|a_{h\rho}|^2\sigma_{tot}^{\eta\rho}=\sigma_{tot}
\end{eqnarray}

The result shows that the model is manifestly consistent with
unitarity, as the total cross section, obtained through the
optical theorem, is explicitly given by the sum of all possible
intermediate states obtained cutting the forward amplitude.

In the gaussian case

\begin{equation}\sigma_{\eta\rho}(b)=\nu_{\eta\rho}(s){{\rm
exp}[-b^2/R^2_{\eta\rho}(s)]\over\pi
R^2_{\eta\rho}(s)}\end{equation}

\noindent the cross
sections are expressed by close analytic forms:

\begin{eqnarray}
\sigma_{el}^{\eta\rho}&=&\pi R^2_{\eta\rho}(s)\Bigl\{2{\rm
Re}\bigl[\gamma+{\rm
ln}\kappa_{\eta\rho}+E_1(\kappa_{\eta\rho})\bigr]
-\bigl[\gamma+{\rm
ln}\kappa'_{\eta\rho}+E_1(\kappa'_{\eta\rho})\bigr]\Bigr\}\cr\nonumber
\sigma_{T}^{\eta\rho}&=&2\pi R^2_{\eta\rho}(s){\rm
Re}\bigl[\gamma+{\rm
ln}\kappa_{\eta\rho}+E_1(\kappa_{\eta\rho})\bigr]\cr
\sigma_{in}^{\eta\rho}&=&\pi R^2_{\eta\rho}(s)\bigl[\gamma+{\rm
ln}\kappa'_{\eta\rho}+E_1(\kappa'_{\eta\rho})\bigr]\nonumber\end{eqnarray}

\par\noindent where $\gamma$ is Euler's constant,

\begin{equation}\kappa_{\eta\rho}={\nu_{\eta\rho}(s)\over \pi R^2_{\eta\rho}(s)}{1-{\rm
i}\alpha_{\eta\rho}\over2},
\qquad\kappa'_{\eta\rho}={\nu_{\eta\rho}(s)\over \pi
R^2_{\eta\rho}(s)}\end{equation}

\par\noindent and $E_1(x)$ the exponential integral.

Explicit expressions for the hadron-hadron total, elastic, single and double diffractive cross
sections at different energies may hence be obtained.

\vskip.25in
\par    {\bf Hard and soft Pomerons }

\vskip.15in

\par\noindent
A cut-Pomeron may be either soft, when there are no large $p_t$
partons in the final state, or hard, when large $p_t$ partons are
present. The two contributions are conveniently discussed with the help of the eigenstates of the forward $T$ matrix. One may write

\begin{equation}\sigma_{\eta\rho}=\sigma_{\eta\rho}^S+\sigma_{\eta\rho}^J\end{equation}

\par\noindent
The contributions to $D_{k,m}^{\eta\rho}$ due to soft
and hard Pomeron exchanges are derived by expanding the
powers of $\sigma_{\eta\rho}$ in Eq.20:

\begin{eqnarray}
D_{k,m}^{\eta\rho}&=&\int d^2b \frac{1}{k!}\binom{k}{ m}
 \biggl[
 \frac{{\rm i}t_{\eta\rho}(b)}{2}+\biggl(\frac{{\rm i}t_{\eta\rho}(b)}{2}\biggr)^*
 \biggr]^{k-m}\times [\sigma_{\eta\rho}^S(b)+\sigma_{\eta\rho}^J(b)]^m\nonumber \\
&=&\sum_{n=0}^m\int d^2b \frac{1}{k!}\binom{k}{ m}\binom{m}{ n}
\bigl( -\sigma_{\eta\rho}(b) \bigr)^{k-m}
\times[\sigma_{\eta\rho}^J(b)]^n[\sigma_{\eta\rho}^S(b)]^{m-n}=\sum_{n=0}^mD_{k,m}^{\eta\rho(n)}
\end{eqnarray}

\par\noindent
The contribution to the inelastic cross section, of the eigenstates of the $T$ matrix $\phi_{\eta}$ and $\phi_{\rho}$, $\sigma_{in}^{\eta\rho}$, with $k$ Pomerons
exchanged and $m$ Pomerons cut ($m\le k$), where one distinguishes
between $n$ hard ($n\le m$) and $m-n$ soft Pomerons, is
represented by $D_{k,m}^{\eta\rho(n)}$.

\par\noindent
The cross section $\sigma_{hard}^{\eta\rho(n)}$, corresponding to
the case of $n$ hard cut-Pomerons, where all soft cut-Pomerons and
all elastic scatterings have been summed, is expressed by

\begin{eqnarray}\sigma_{hard}^{\eta\rho(n)}=\sum_{k=n}^{\infty}\sum_{m=n}^kD_{k,m}^{\eta\rho}&=&\int d^2b\sum_{k=n}^{\infty} \frac{1}{ n!(k-n)!}
                        [\sigma_{\eta\rho}^J(b)]^n\bigl[\sigma_{\eta\rho}^S(b)-\sigma_{\eta\rho}(b)\bigr]^{k-n}\cr
                        &=&\int d^2b\frac{1}{ n!}\bigl(\sigma_{\eta\rho}^J(b)\bigr)^n
    {\rm e}^{-\sigma_{\eta\rho}^J(b)} \end{eqnarray}

\par\noindent
the cross section $\sigma_{hard}^{\eta\rho}$, including all events
with hard cut-Pomerons, is obtained after summing all terms
$\sigma_{hard}^{\eta\rho(n)}$ with $n>0$:

\begin{equation}\sigma_{hard}^{\eta\rho}=\sum_{n=1}^{\infty}\int d^2b\frac{1}{ n!}\bigl(\sigma_{\eta\rho}^J(b)\bigr)^n
    {\rm e}^{-\sigma_{\eta\rho}^J(b)}=\int d^2b\Bigl\{1-
    {\rm e}^{-\sigma_{\eta\rho}^J(b)}\Bigr\}\end{equation}

\par\noindent
The cross sections $\sigma_{tot}^{\eta\rho}$, $\sigma_{in}^{\eta\rho}$
and $\sigma_{hard}^{\eta\rho}$ are all expressed by Poissonians at
fixed impact parameter: $\sigma_{tot}^{\eta\rho}$ is a
superposition of uncut-Pomeron amplitudes,
$\sigma_{in}^{\eta\rho}$ is a superposition of cut-Pomerons and
$\sigma_{hard}^{\eta\rho}$ is a superposition of hard
cut-Pomerons. In particular only hard processes contribute to the
shadowing corrections of $\sigma_{hard}^{\eta\rho}$, which hence
belongs to the category of the self-shadowing cross
sections\cite{Capella:1981ju}.

\vskip .25in
\par   {\bf Soft cross section}
\vskip .15in

\par\noindent Let's consider $D_{k,m}^{{\eta\rho}(0)}$,
namely the contribution to the inelastic cross section of the states $\phi_{\eta}$ and $\phi_{\rho}$ with $k$
Pomerons exchanged and $m$ soft cut-Pomeron ($m\le k$)

\begin{equation}D_{k,m}^{{\eta\rho}(0)}=\int d^2b \frac{1}{k!}\binom{k}{ m}
                      \bigl[-\sigma_{\eta\rho}(b)\bigr]^{k-m}
                      \bigl[\sigma_{\eta\rho}^S(b)\bigr]^{m}\end{equation}

\par\noindent The contribution to the soft inelastic cross section, with $m$ soft cut-Pomeron $\sigma_m^{{\eta\rho}(0)}$, is obtained summing all
elastic scatterings in $D_{k,m}^{{\eta\rho}(0)}$:

\begin{eqnarray}\sigma_m^{{\eta\rho}(0)}&=&\sum_{k=m}^{\infty}D_{k,m}^{{\eta\rho}(0)}=\int d^2b \sum_{k=m}^{\infty}
\frac{1}{
m!(k-m)!}\bigr[-\sigma_{\eta\rho}(b)\bigl]^{k-m}\bigl[\sigma_{\eta\rho}^S(b)\bigr]^m\cr
&=&\int d^2b
\frac{1}{m!}\bigl[\sigma_{\eta\rho}^S(b)\bigr]^m\sum_{l=0}^{\infty}
\frac{\bigr[-\sigma_{\eta\rho}(b)\bigl]^l}{ l!}\cr &=&\int d^2b
\frac{1}{m!}\bigl[\sigma_{\eta\rho}^S(b)\bigr]^m{\rm e}
^{-\sigma_{\eta\rho}(b)} \end{eqnarray}

\par\noindent The soft cross section $\sigma_{soft}^{\eta\rho}$, including all soft cut-Pomerons and all additional elastic scatterings, is hence given by

\begin{eqnarray}\sigma_{soft}^{\eta\rho}&=&\sum_{m=1}^{\infty}\sigma_m^{{\eta\rho}(0)}\cr&=&\int d^2b
\Bigl\{{\rm e} ^{\sigma_{\eta\rho}^S(b)-\sigma_{\eta\rho}(b)}-{\rm
e} ^{-\sigma_{\eta\rho}(b)}\Bigr\}\cr&=&\int d^2b \Bigl\{{\rm e}
^{-\sigma_{\eta\rho}^J(b)}-{\rm e}
^{-\sigma_{\eta\rho}(b)}\Bigr\}\cr
&=&\quad\sigma_{in}^{\eta\rho}-\sigma_{hard}^{\eta\rho}
\end{eqnarray}

\par\noindent The sum of all intermediate states, generated by cut-Pomerons between eigenstates of the $T$ matrix, gives the inelastic cross section; the sum over all
intermediate states, generated without cutting Pomerons, gives the
elastic cross section and the sum of the two different
contributions gives the total cross section.
\par\noindent When considering hard cut-Pomerons, the sum over all
intermediate states gives the hard cross section; the sum over all
intermediate states, generated without hard cut-Pomerons, gives
the soft cross section plus the elastic cross section and the sum
of the two different contributions gives the total cross section.

The hard and soft cross sections between the hadronic states $\mu$
and $\nu$ are therefore

\begin{eqnarray}
(\sigma_{hard})_{\mu\nu}&=&\sum_{\eta\rho}|a_{h\eta}|^2|a_{h\rho}|^2\sigma_{hard}^{\eta\rho}\cr
(\sigma_{soft})_{\mu\nu}&=&\sum_{\eta\rho}|a_{h\eta}|^2|a_{h\rho}|^2\sigma_{soft}^{\eta\rho}=(\sigma_{in})_{\mu\nu}-(\sigma_{hard})_{\mu\nu}
\end{eqnarray}

\section{Average, dispersion and inclusive cross sections }

Let's now go to the case of interest, where the initial hadron
state is the proton. The indices $\mu$, $\nu$ may hence be
dropped. The inelastic $pp$ cross sections is

\begin{eqnarray}\sigma_{in}=
\sum_{\eta\rho}|a_{\eta}|^2|a_{\rho}|^2\sigma_{in}^{\eta\rho}&=&
\sum_{\eta\rho}|a_{\eta}|^2|a_{\rho}|^2 \int
d^2b\sum_{m=1}^{\infty}\frac{1}{m!}\bigl[\sigma_{\eta\rho}(b)\bigr]^m
                            {\rm
e}^{-\sigma_{\eta\rho}(b)}\cr &=&
\sum_{\eta\rho}|a_{\eta}|^2|a_{\rho}|^2 \int d^2b\Bigl\{1-{\rm
e}^{-\sigma_{\eta\rho}(b)}\Bigr\}\end{eqnarray}

\noindent and the hard cross section is analogously expressed by

 \begin{eqnarray}\sigma_{hard}=
\sum_{\eta\rho}|a_{\eta}|^2|a_{\rho}|^2\sigma_{hard}^{\eta\rho}&=&
\sum_{\eta\rho}|a_{\eta}|^2|a_{\rho}|^2 \int
d^2b\sum_{m=1}^{\infty}\frac{1}{m!}\bigl[\sigma_{\eta\rho}^J(b)\bigr]^m
                            {\rm
e}^{-\sigma_{\eta\rho}^J(b)}\cr &=&
\sum_{\eta\rho}|a_{\eta}|^2|a_{\rho}|^2 \int d^2b\Bigl\{1-{\rm
e}^{-\sigma_{\eta\rho}^J(b)}\Bigr\}\end{eqnarray}

\noindent while the soft cross section is given by the difference

\begin{equation}\sigma_{soft}=\sigma_{in}-\sigma_{hard}=\sum_{\eta\rho}|a_{\eta}|^2|a_{\rho}|^2 \int
d^2b\Bigl\{{\rm
e}^{-\sigma_{\eta\rho}^J(b)}-{\rm
e}^{-\sigma_{\eta\rho}(b)}\Bigr\}\end{equation}

\noindent which shows that the soft cross section is a border
effect and goes to zero in the high energy-fixed momentum exchange
limit.

The average number of inelastic collisions is given by the single Pomeron exchange term between the different eigenstates of the $T$ matrix

\begin{eqnarray}\langle m\rangle\sigma_{in}&=&
\sum_{\eta\rho}|a_{\eta}|^2|a_{\rho}|^2 \int
d^2b\sum_{m=1}^{\infty}\frac{m}{m!}\bigl[\sigma_{\eta\rho}(b)\bigr]^m
                            {\rm
e}^{-\sigma_{\eta\rho}(b)}\cr &=&
\sum_{\eta\rho}|a_{\eta}|^2|a_{\rho}|^2 \int
d^2b\sigma_{\eta\rho}(b)=\sum_{\eta\rho}|a_{\eta}|^2|a_{\rho}|^2
\nu_{\eta\rho}(s)\end{eqnarray}

\noindent and analogously the average number of hard collisions is
given by the single  scattering inclusive cross section
of the perturbative QCD parton model

\begin{eqnarray}\langle m\rangle\sigma_{hard}&=&
\sum_{\eta\rho}|a_{\eta}|^2|a_{\rho}|^2 \int
d^2b\sum_{m=1}^{\infty}\frac{m}{m!}\bigl[\sigma_{\eta\rho}^J(b)\bigr]^m
                            {\rm
e}^{-\sigma_{\eta\rho}^J(b)}\cr &=&
\sum_{\eta\rho}|a_{\eta}|^2|a_{\rho}|^2 \int
d^2b\sigma_{\eta\rho}^J(b)\cr
&=&\sigma_{QCD}(s,p_{cut})|_{single}\end{eqnarray}

\noindent The double scattering inclusive cross sections are
given by the dispersion of the multiplicity
distribution in the number of collisions:

\begin{eqnarray}{1\over2}\langle m(m-1)\rangle\sigma_{in}&=&
{1\over2}\sum_{\eta\rho}|a_{\eta}|^2|a_{\rho}|^2 \int
d^2b\sum_{m=1}^{\infty}\frac{m(m-1)}{m!}\bigl[\sigma_{\eta\rho}(b)\bigr]^m
                            {\rm
e}^{-\sigma_{\eta\rho}(b)}\cr
&=&{1\over2}\sum_{\eta\rho}|a_{\eta}|^2|a_{\rho}|^2 \int
d^2b\bigl[\sigma_{\eta\rho}(b)\bigr]^2={1\over2}\sum_{\eta\rho}|a_{\eta}|^2|a_{\rho}|^2
{\nu_{\eta\rho}^2(s)\over2\pi R^2_{\eta\rho}(s)}\end{eqnarray}

\noindent and

\begin{eqnarray}{1\over2}\langle m(m-1)\rangle\sigma_{hard}&=&{1\over2}\sum_{\eta\rho}|a_{\eta}|^2|a_{\rho}|^2 \int
d^2b\sum_{m=1}^{\infty}\frac{m(m-1)}{m!}\bigl[\sigma_{\eta\rho}^J(b)\bigr]^m
                            {\rm
e}^{-\sigma_{\eta\rho}^J(b)}\cr
&=&{1\over2}\sum_{\eta\rho}|a_{\eta}|^2|a_{\rho}|^2 \int
d^2b\bigl[\sigma_{\eta\rho}^J(b)\bigr]^2\cr
&=&\sigma_{QCD}(s,p_{cut})|_{double}={1\over2}
{\bigl[\sigma_{QCD}(s,p_{cut})|_{single}\bigr]^2\over\sigma_{eff}}\end{eqnarray}

\noindent where the scale factor $\sigma_{eff}$, characterizing
double parton interaction processes has been introduced. \noindent
An analogous scale factor may be introduced for double Pomeron exchanges:

\begin{equation}\sigma_{eff, P}\equiv{\Bigl[\sum_{\eta\rho}|a_{\eta}|^2|a_{\rho}|^2 \nu_{\eta\rho}(s)\Bigr]^2\over \sum_{\eta\rho}|a_{\eta}|^2|a_{\rho}|^2 {\nu_{\eta\rho}^2(s)\over2\pi
R^2_{\eta\rho}(s)}}\end{equation}

\noindent Introducing the couplings of the Pomeron to the
eigenstate of the T matrix ${\eta}$ and ${\rho}$, $g_{\eta}$ and
$g_{\rho}$, and the Pomeron trajectory

\begin{equation}\alpha_P(t)=1+\Delta_P+\alpha_P' t
\end{equation}

\noindent one has

\begin{equation}\nu_{\eta\rho}(s)=g_{\eta}g_{\rho}\Bigl({s\over s_0}\Bigr)^{\Delta_P}\end{equation}

\noindent and the scale factor for double Pomeron exchanges, $\sigma_{eff, P}$,
is given by

\begin{equation}\sigma_{eff, P}\equiv
{\Bigl[\sum_{\eta\rho}|a_{\mu\eta}|^2|a_{\nu\rho}|^2
g_{\eta}g_{\rho}\Bigr]^2\over
\sum_{\eta\rho}|a_{\mu\eta}|^2|a_{\nu\rho}|^2
{g_{\eta}^2g_{\rho}^2\over2\pi R^2_{\eta\rho}(s)}}\end{equation}

The scale factor $\sigma_{eff, P}$ has a very smooth energy
dependence, in fact it depends on $s$ only logarithmically
(through $R_{\eta\rho}(s)$). Similar properties hold for
$\sigma_{eff}$ which, although related to a hard process, is a
cutoff independent quantity. In the high energy (and fixed cutoff)
limit the inelastic cross section will be saturated by the hard
cross section. One may hence argue that, in the regime were the
soft component of the interaction becomes negligible,
$\sigma_{eff, P}$ and $\sigma_{eff}$ will coincide. $\sigma_{eff}$
however should not depend on energy and cutoff (apart from the
logarithmic dependence of the hadron radius on $s$) which brings
to the conclusion that if $\sigma_{eff, P}$ and $\sigma_{eff}$
coincide in the regime were the soft component of the interaction
becomes negligible, the two effective cross sections will coincide
at any regime.

\vskip .25in
\par   {\bf The simplest two channel case}
\vskip .15in

\par

The simplest two channel case has been studied in detail in the
literature and it was shown to be able to reproduce with accuracy
available data of total, elastic, single and double diffraction
cross sections of $pp$ collisions at different
energies\cite{Gotsman:1999ri}\cite{Gotsman:2000gb}\cite{Gotsman:2007pn}.
In the two channel formalism the states observed
are either the initial hadron ($\psi_h$) or its diffractive
states, represented globally by a single channel ($\psi_D$). The
eigenstates of the forward interaction operator $T$ are $\phi_1$
and $\phi_2$. One may hence write

\begin{equation}\psi_h=\alpha\phi_1+\beta\phi_2\end{equation}

\noindent while $\psi_D$ is represented by the orthogonal
superposition

\begin{equation}\psi_D=-\beta\phi_1+\alpha\phi_2\end{equation}

\noindent with $\alpha^2+\beta^2=1$ (hence the relation with the
coefficients $a_{\mu\eta}$ introduced in the previous paragraphs
is $a_{11}=a_{22}=\alpha$, $a_{12}=-a_{21}=\beta$).

\noindent The total and inelastic hadron-hadron cross sections are

\begin{equation}\sigma_{tot}=\sigma_{tot}^{11}\alpha^4+\sigma_{tot}^{12}\alpha^2\beta^2+\sigma_{tot}^{21}\alpha^2\beta^2+\sigma_{tot}^{22}\beta^4\end{equation}

\begin{equation}\sigma_{in}=\sigma_{in}^{11}\alpha^4+\sigma_{in}^{12}\alpha^2\beta^2+\sigma_{in}^{21}\alpha^2\beta^2+\sigma_{in}^{22}\beta^4\end{equation}

\noindent and

\begin{eqnarray}\sigma_{el}+\sigma_{SD}+\sigma_{DD}=\sigma_{el}^{11}\alpha^4+2\sigma_{el}^{12}(\alpha\beta)^2+\sigma_{el}^{22}\beta^4\end{eqnarray}

\noindent Unitarity is satisfied, as it may be checked summing all
contributions

\begin{eqnarray}\sigma_{el}+\sigma_{SD}+\sigma_{DD}+\sigma_{in}&=&\sigma_{el}^{11}\alpha^4+2\sigma_{el}^{12}(\alpha\beta)^2+
\sigma_{el}^{22}\beta^4\cr&&+\sigma_{in}^{11}\alpha^4+\sigma_{in}^{12}\alpha^2\beta^2+\sigma_{in}^{21}\alpha^2\beta^2+\sigma_{in}^{22}\beta^4\cr
&=&\sigma_{tot}^{11}\alpha^4+\sigma_{tot}^{12}\alpha^2\beta^2+\sigma_{tot}^{21}\alpha^2\beta^2+\sigma_{tot}^{22}\beta^4=\sigma_{tot}\end{eqnarray}

The most appropriate choice of parameters for reproducing the
total, elastic, single and double diffraction cross sections is
the following\cite{Gotsman:2007pn}

\begin{eqnarray}
\nonumber
\Delta_P&=&0.15,\quad\alpha^{'}=0.173GeV^{-2},\quad s_0=1GeV^{2},\\
\nonumber \sigma^{0}_{11}&=&9.22GeV^{-2},\quad
\sigma^{0}_{22}=3503.5GeV^{-2},\quad
\sigma^{0}_{12}=6.53GeV^{-2},\\
\nonumber \sigma^{0}_{i,k}&=&g_{i}g_{k},\quad\beta=0.776,\quad
R_{0,1}^2=10.42GeV^{-2},\quad r_0^{2}=0.5GeV^{-2}\end{eqnarray}

\noindent where

\begin{eqnarray}\nonumber
R_{11}^2(s)&=&2R_{0,1}^2+r_0^2+4\alpha'{\rm
ln}(s/s_0)\\
\nonumber
R_{12}^2(s)&=&R_{0,1}^2+r_0^2+4\alpha'{\rm
ln}(s/s_0)\\
\nonumber R_{22}^2(s)&=&r_0^2+4\alpha'{\rm ln}(s/s_0)
\end{eqnarray}

\noindent With this input one obtains

\begin{eqnarray}
\nonumber \sqrt(s)&=&14{\rm TeV}\qquad\sigma_{tot}=114{\rm
mb}\qquad\sigma_{inel}=71{\rm mb}\qquad\sigma_{eff}=12{\rm
mb}\\
\nonumber \sqrt(s)&=&1.8{\rm TeV}\qquad\sigma_{tot}=81{\rm
mb}\qquad\sigma_{inel}=50{\rm mb}\qquad\sigma_{eff}=10{\rm
mb}\qquad
\end{eqnarray}

\noindent where $\sigma_{eff}$ has been identified with
$\sigma_{eff, P}$ as expressed in eq. 54.

Notice that in a single channel gaussian model the value of the
effective cross section is given by $2\pi R^2$ (cfr eq.54) with
$R$ the radius of the overlap function of the matter distribution
of the two hadrons, corresponding to $R^2=4/3\langle r^2\rangle$,
where $\sqrt\langle r^2\rangle$ is the rms hadron radius. For the
proton one might take $\sqrt\langle r^2\rangle=.6$
fm\cite{Abe:1997xk}. The effective cross section would then be of
the order of $30$ mb, much larger with respect to the experimental
indications. In the multi-channel approach, on the contrary, one
obtains that the effective cross section results from the sum in
eq.52, which leads necessarily to a smaller value with respect to
the single channel case. The weights of the different channels are
related to the value of the diffractive cross section.
Interestingly the simplest multichannel model, able to reproduces
the observed values of the diffractive cross section, leads to an
effective cross section which, as discussed in the next paragraph,
compares well with the experimental indications.

\vskip .25in
\par   {\bf Effective cross section}
\vskip .15in

\par
In the analysis of double parton collisions the CDF collaboration has
claimed that the measured value of the effective cross section is
consistent with the result of the simplest single channel picture,
where the hadron matter density is characterized by the size of
the conventional rms hadron radius\cite{Abe:1997bp}\cite{Abe:1997xk}. The claim is the consequence
of an unfortunate mistake (in eq. 11 of ref.\cite{Abe:1997xk}
there shouldn't be any factor $1/2$) which affects the conclusions
of CDF concerning the importance of parton correlations in the
process, although it does not affect the actual value of the scale factor
measured in the experiment.

The effective cross section quoted by
CDF, $\sigma_{eff}=14.5\pm1.7^{+1.7}_{-2.3}$, is nevertheless different
with respect to the effective cross section discussed here above
and in most of the papers on double parton scatterings.
$\sigma_{eff}$ has in fact a simple link with the overlap of
matter distribution in the hadronic collision only when, as
previously explained, the double scattering cross section is obtained
from the dispersion of the distribution in the number of
partonic collisions. In the analysis of CDF all events with triple
parton scatterings have on the contrary been removed from the
sample of events with double parton collisions. The experiment in
fact has measured the {\it exclusive} rather than the {\it
inclusive} double parton scattering cross section, namely it has
measured the contribution to the total inelastic cross section due
to double parton collisions. As in the case of CDF the fraction of
events with triple scatterings is 17\% of the collected sample,
the difference between the two quantities is not negligible.

\par
An indication on the actual value of the scale factor may be
obtained making a few simplifying hypotheses. The experiment has
measured the rate of events with three minijets and one prompt
photon. One may neglect the contamination of events with four or
more parton collisions. Let's call ${\rm P}_2^{A,B}(\beta)$ and
${\rm P}_3^{A,B}(\beta)$ the probabilities of double and triple
parton scattering at fixed impact parameter, the two differnt
parton processes being the parton collision giving a photon + a
minijet ($A$) and two minijets ($B$) respectively. Let's in
addition assume that triple scatterings are only due to single
collisions of kind $A$ and double collisions of kind $B$.

\noindent Since the contamination of the collected sample due to
triple collisions is $17\%$ one may estimate:

\begin{eqnarray}
\sigma_D^{A,B}&=&\langle N_B\rangle\sigma_{hard}^{A,B}\nonumber\\
&\simeq&\int d^2\beta\quad {\rm
P}_2^{A,B}(\beta)+2\int
d^2\beta\quad {\rm P}_3^{A,B}(\beta)\nonumber\\
&=&\bigl[\sigma_D\bigr]_{CDF}+2\times\frac{17}{
83}\bigl[\sigma_D\bigr]_{CDF}\approx1.34\bigl[\sigma_D\bigr]_{CDF}
\end{eqnarray}

\par\noindent
where the factor 2 in front of ${\rm P}_3^{A,B}(\beta)$ is due to
the multiplicity of collisions of kind $B$. One hence obtains

\begin{equation}
\sigma_{eff}=\frac{\bigl(\sigma_{eff}\bigr)_{CDF}}{1.34}\approx\quad
11{\rm mb}
\end{equation}

Amazingly the result is rather close to the value obtained in the
multichannel eikonal model.

\section{Concluding remarks}

The aim of this paper is to study the relation between multiple
parton interactions and hadronic diffraction. To that purpose the
multichannel eikonal model of high energy hadronic interactions
has been discussed in detail, establishing an explicit relation
between the two phenomena. The puzzling feature of multiparton
interactions, the small value of the effective cross section,
finds a natural explanation in this framework. From a qualitative
point of view the reduction of the value of the effective cross
section, with respect to the value obtained in a single channel
model of high energy hadronic interactions, is an intrinsic
feature of the multichannel approach. Interestingly, the simplest
eikonal model, able to reproduce high energy hadronic diffraction,
gives a value of the effective cross close to the experimental
indications.

The non perturbative input of the multiparton cross
section is represented by the multiparton distribution functions,
which depend on the scale of the process, on the kinematical
variables, on the different kinds of interacting partons and on
their relative transverse distance. In a multiparton collision,
the relative transverse distance is naively related to the hadron
form factor, which leads however to an effective cross section too
large in comparison with experiment. The main result of the
present analysis is that the form factor is an average over many
different hadronic configurations, which however interact
with different strengths in a high energy hadronic collisions.
Multiple parton interactions are more likely to take place when
the hadron is in a compact configuration, which enhances
relatively small transverse distances, leading to relatively small
values of the effective cross section.

\section{Acknowledgment}I thank Uri Maor for many enlightening discussions.

\end{document}